\pdfoutput=1

\documentclass[11pt]{article}

\usepackage[final]{acl}

\usepackage{times}
\usepackage{latexsym}

\usepackage[T1]{fontenc}

\usepackage[utf8]{inputenc}

\usepackage{microtype}

\usepackage{inconsolata}

\usepackage{graphicx}

\usepackage[ruled]{algorithm2e}
\usepackage{algorithmic}
\usepackage{booktabs}

\usepackage{subfig}
\usepackage{arydshln}
\usepackage{multirow}
\usepackage{multicol}
\usepackage{amsmath}
\usepackage{amssymb}
\usepackage{colortbl}
\usepackage{tabu}                     
\usepackage{multirow}                 
\usepackage{multicol}                 
\usepackage{multirow}                
\usepackage{float}                    
\usepackage{makecell}                 
\usepackage{booktabs}                 
\definecolor{mygray}{gray}{.85}

\title{Empowering LLMs in Task-Oriented Dialogues: A Domain-Independent Multi-Agent Framework and Fine-Tuning Strategy}

\author{Zihao Feng$^{1,2}$\thanks{Equal contribution}\thanks{Zihao Feng was an intern at Tencent during the preparation of this work}, Xiaoxue Wang$^{2 *}$, Bowen Wu$^{3,2}$, Weihong Zhong$^{1}$, Zhen Xu$^{2}$, \\ \textbf{Hailong Cao$^{1}$, Tiejun Zhao$^{1}$\thanks{Corresponding author}, Ying Li$^{3}$, Baoxun Wang$^{2}$} \\
  $^1$Faculty of Computing, Harbin Institute of Technology \\
  $^2$Platform and Content Group, Tencent \\
  $^3$School of Software \& Microelectronics, Peking University \\
  \texttt{21b903052@stu.hit.edu.cn, whzhong@ir.hit.edu.cn} \\
  \texttt{\{caohailong, tjzhao\}@hit.edu.cn} \\
  \texttt{\{yukixxwang, zenxu, asulewang\}@tencent.com,  \{jason\_wbw, li.ying\}@pku.edu.cn} \\ 
}

\begin{document}
\maketitle
\begin{abstract}

Task-oriented dialogue systems based on Large Language Models (LLMs) have gained increasing attention across various industries and achieved significant results. Current approaches condense complex procedural workflows into a single agent to achieve satisfactory performance on large-scale LLMs. However, these approaches face challenges to achieve comparable performance on fine-tuned lightweight LLMs, due to their limited capabilities in handling multiple complex logic. In this work, we design a Domain-Independent Multi-Agent Framework (DIMF), which contains Intent Classification Agent, Slot Filling Agent and Response Agent. This approach simplifies the learning complexity and enhances the generalization ability by separating the tasks into domain-independent components. In this framework, we enhance the capabilities in contextual understanding using the Direct Preference Optimisation (DPO) method, and propose a simple and effective Data Distribution Adaptation (DDA) method to mitigate degradation issues during DPO training. Experiments conducted on the MultiWOZ datasets show that our proposed method achieves a better average performance among all the baselines. Extensive analysis also demonstrates that our proposed framework exhibits excellent generalizability and zero-shot capability.
\end{abstract}

\section{Introduction}

\begin{figure}[htbp]  
\centering
\includegraphics[width=0.48\textwidth]{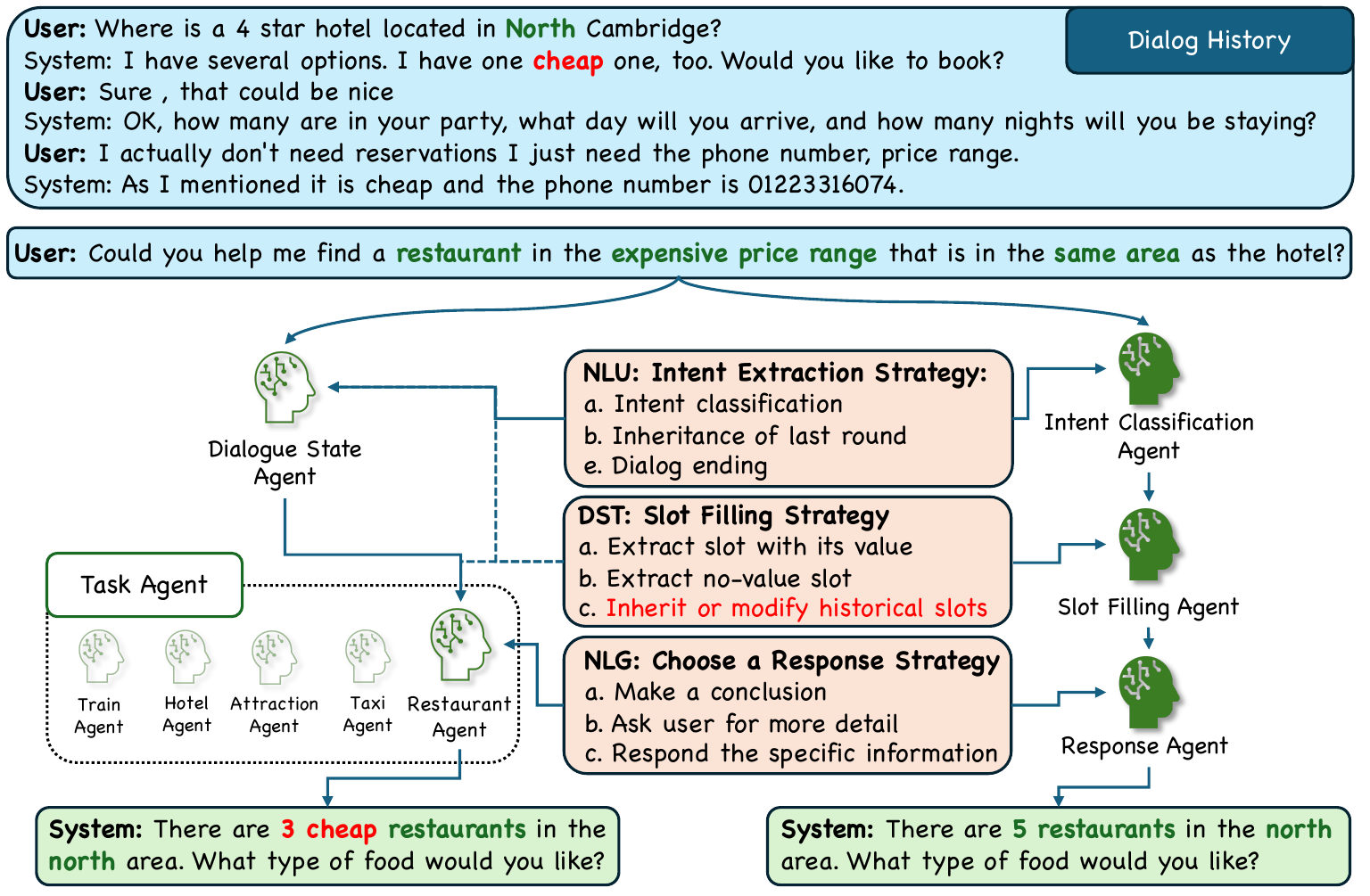} 
\caption{Different architectures of our proposed system and other LLM-based systems. The left part is other LLM-based systems and the right is ours. The information in the orange box indicates the strategies in different sub-tasks that the agent needs to follow. 
} 
\label{introfig}
\end{figure}

Task-oriented dialogue (TOD) systems play a significant role in both academic research and industry.\cite{peng2022godel,xu2024rethinking}. Researchers have divided the traditional TOD systems into the following several key components \cite{zhang2020recent}: 1) Natural Language Understanding (NLU) \cite{karanikolas2023large}. 2) Dialogue State Tracking (DST) \cite{feng2023towards, heck2023chatgpt, feng2025improving}. 3) Dialogue Policy. 4) Natural Language Generation (NLG) \cite{li-etal-2020-slot}. With the development of the Large Language Model (LLM), recent research has mainly focused on leveraging the strong capabilities and generalization of LLMs to solve the complex task of TOD \cite{qin2023end, algherairy2024review, chung2023instructtods}. The LLM-based multi-agent approach has been proven to be effective in multi-domain TOD systems \cite{gupta2024dard}. 

Existing methodologies often attempt to condense complex procedural workflows of TOD systems into a single large-scale LLM-based agent such as GPT-4 \cite{achiam2023gpt} and Claude, or divide the workflow into different domains to conduct multi-agent TOD systems for lightweight LLMs. Most works have achieved satisfactory performance on large-scale LLMs \cite{xu2024rethinking,gupta2024dard}. In contrast, the lightweight models, even when fine-tuned for specific tasks, struggle to attain comparable completion quality \cite{xu2024rethinking,gupta2024dard}.
This discrepancy contrasts sharply with their competitive performance in other NLP tasks 
, suggesting that the inherent complexity of TOD necessitates specialized approaches.
We posit that effective modeling of multi-step procedural logic and developing targeted learning strategies are critical to bridging this performance gap.

To address this challenge, we propose a Domain-Independent Multi-Agent Framework (DIMF), which contains Intent Classification Agent, Slot Filling Agent and Response Agent.
Unlike the current methods, which conduct multi-agent system by different domain-specific agents, DIMF decouples the workflow into several components which are domain-independent.
As illustrated in Figure \ref{introfig}, both phases require contextual reasoning and policy-guided decision-making capabilities, easily conflated in monolithic agent architectures. 
The task separation design stems from our observation of domain relevance and challenges in slot integration from dialogue history during slot filling process. This approach guarantees that the agent considers the slot that matches the current specific domain.
Furthermore, this modular decomposition facilitates the enhancement of targeted capability through reinforcement learning techniques (e.g., DPO/PPO \cite{rafailov2023direct, schulman2017proximal}), enabling specialized optimization while maintaining domain adaptability. 
We therefore propose a Data Distribution Adaptation (DDA) method designed to mitigate the degradation of DPO training attributable to the diversity of domain types.




The experimental results indicate that the framework and training methodology significantly enhance the performance of the fine-tuned models. 
Additionally, it was observed that the domain-independent design exhibits a robust zero-shot capability. 
In conclusion, this paper offers the following contributions:


\begin{itemize}
    \item We design a novel Domain-Independent Multi-Agent Framework for TOD systems based on LLMs. Our approach separates the complex task into three sub-tasks which better leverages the generalization capabilities of LLMs.
    \item We utilize DPO during the training process, and innovatively propose a Data Distribution Adaptation method to alleviate the DPO's training degradation problem during the DPO training process.  
    \item Our new framework and training strategy for the TOD system have enhanced the system's scalability and zero-shot capabilities, allowing the system to maintain good performance even on domains it has not seen before. 
\end{itemize}

\section{Background}

\subsection{Large Language Models as Agents}
Recently, many efforts have been made to build systems through LLMs acting as agents for planning, decision-making, and acting tasks between various specialized APIs, dialogue, or other simpler tools to perform complex tasks \cite{liu2023agentbench, liang2023encouraging, deng2024mind2web}. ReAct \cite{yao2023react} method is a prompt framework that has been widely used for fine-tuning the LLMs with the ability of reasoning and action based on text. Various tasks such as logical reasoning \cite{du2023improving, tang2023medagents}, societal simulations \cite{zhou2023sotopia}, tool learning \cite{qin2023toolllm, shen2024small} have achieved significant improvement in performance using LLMs as agents.

However, most research focuses on task-specific scenarios with poor scalability. The challenge of LLMs working as agents that can generalize better and adapt to different tasks needs more research.

\subsection{Direct Preference Optimisation (DPO)}
\label{Back_DPO}
Direct Preference Optimisation (DPO) \cite{rafailov2024direct} is a popular method for learning from human-preference data, and it has been widely leveraged to improve the performance of pre-trained LLMs on downstream tasks 
\cite{wang2023making, tunstall2023zephyr}. DPO directly uses pair-wise preference data for model optimization. In this way, we can directly train the language model through the reward learning pipeline, eliminating the need for the reinforcement learning stage.

\begin{figure*}[htbp]  
\centering
\includegraphics[width=\textwidth]{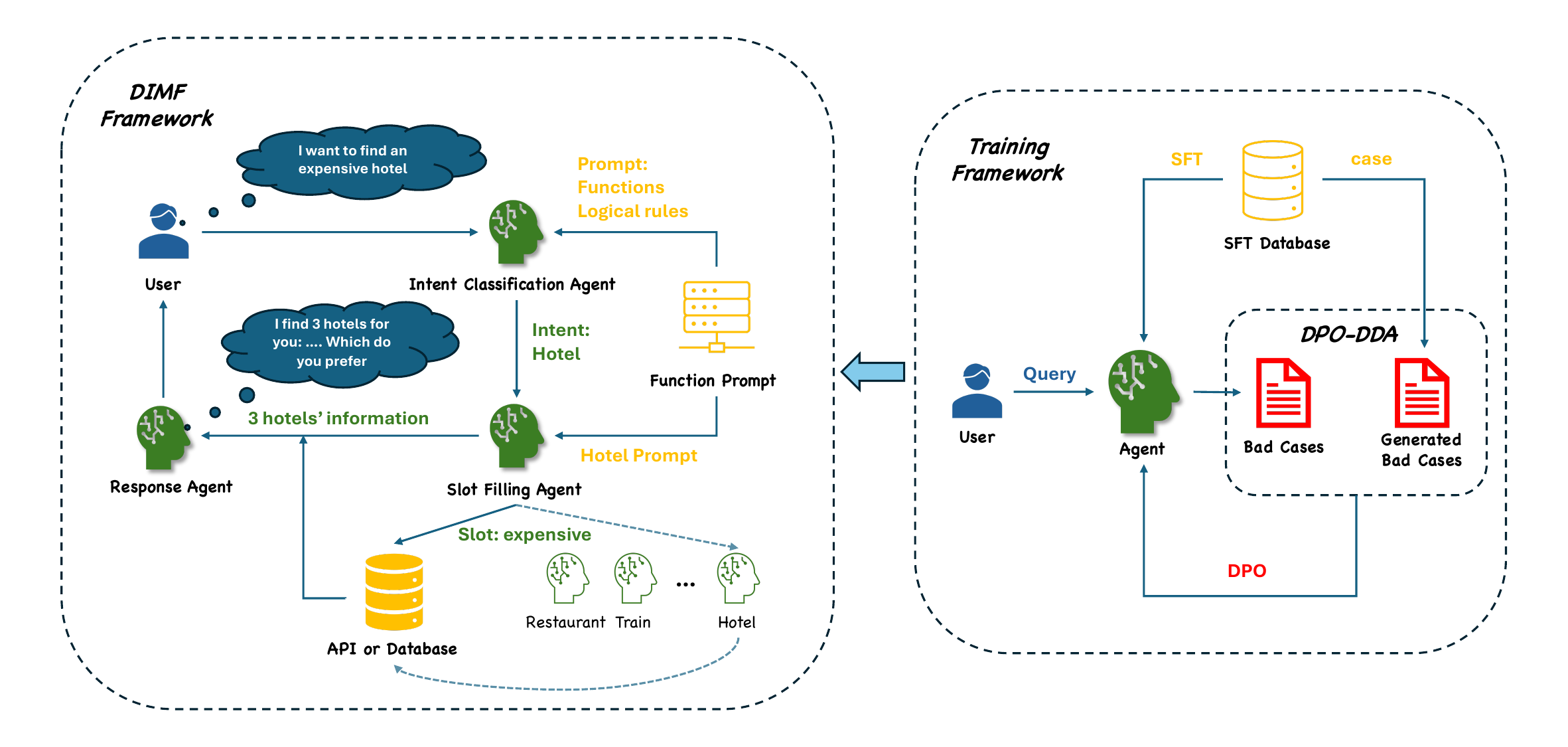} 
\caption{The main framework of our proposed method. 
The left part is the framework of our proposed DIMF. We train three agents to collaboratively solve users' questions and provide responses. Each agent can fulfill different user needs through different prompts, instead of training domain-specific agents (as indicated by the agents in the left part such as "Restaurant"). The right part is the framework of our training process for each agent. We first fine-tune the model with the training set, and then leverage the validation dataset to complete the DPO process.} 
\label{framework}
\end{figure*}

Although the DPO method facilitates model training, experiments demonstrate that the 
DPO loss has flaws: Compared to learning to generate responses preferred by humans, the DPO loss function demonstrates a tendency for LLMs to readily learn to avoid generating responses that humans disprefer \cite{feng2024towards}. Based on this conclusion, DPO exhibits significant degradation issues on data where the Levenshtein Distance between positive and negative examples is small. The reason is that with highly similar positive and negative examples, the DPO process tends to reject the negative examples, which in turn reduces the generation probability for the corresponding positive examples \cite{pal2024smaug}. Thus, the DPO process can lead to a simultaneous decrease in the reward functions for both positive and negative examples, which leads to degradation.

\section{Domain-Independent Multi-Agent Framework}
In this section, we introduce our proposed Domain-Independent Multi-Agent Framework (DIMF) for the TOD task. We give an introduction to the Intent Classification Agent
, Slot Filling Agent 
and Response Agent 
separately. We will provide a detailed introduction to the division of labor 
between each agent.

\subsection{Intent Classification Agent}
\label{IA}
The Intent Classification Agent 
aims to extract the intent of the user's question and serves as the foundation for the subsequent agents. 
Specifically, this agent is provided with the user's question and the descriptions of each domain, then outputs in the ReACT format. 
Besides, this task involves the user's follow-up questions regarding historical dialogue. Therefore, we have designed a logic module in the prompt that provides the logical rules in the current round of dialogue based on the intent of the last round.
Moreover, we design an "other" domain to implement the dialogue-ending intent
. The details of the prompt are appended in Appendix \ref{IAprompt}. 

\subsection{Slot Filling Agent}
\label{SA}
After obtaining the intent of the user's question from the Intent Classification Agent, we train a Slot Filling Agent to extract slots for the specific domain from the query, which is required for extracting information from the database. 
This agent can be adapted to various domains through conducting domain-specific prompts. In this way, we can obtain a generalized Slot Filling Agent instead of training different models for different domains.

For the user's questions, there are two different types of slots: 1) The slot with its corresponding value, such as \emph{I need train reservations from Norwich to Cambridge.} which contains the name of the departure and destination. 2) The slot without value, such as \emph{I would also like to know the travel time, price, and departure time please.} which needs to respond the value to the user. We design two modules to respond to these two types of information separately, and provide a logical rules module in the prompt to distinguish between them. 

Besides, to address the issue of slot inheritance based on dialogue history, we have also designed a module for the Slot Filling Agent in the prompt that includes historical dialogue slots, allowing the agent to better implement this capability by integrating this information with the dialogue history.
Later, according to the generated slot information by the Slot Filling Agent, we can extract the entries in the database that match the user's query. In this work, we use a rule-based approach for extraction.
The detail of the prompt is attached in the Appendix \ref{SAprompt}. 

\subsection{Response Agent}
\label{RA}
Different dialogue histories and states dictate various strategies, such as asking the user to fill in the required slots, allowing the user to refine results, letting the user confirm or cancel, and so on. The Response Agent aims to respond to the user based on the dialogue history and states. Since the database's results of each query vary, we develop the following strategies for the Response Agent to assist the user in obtaining information about the outcome during conversations.

After calling database, the response strategy depends on the number of database results that meet the user's question. If there is only one option, the agent should respond to the information of a specific item that the user asks directly. Otherwise, the response's content should contain the following information: 1) The total number of available options. 2) The conclusion of all options. 3) The question asking users for more specific information to narrow the range of available options. 
The detail of the prompt is attached in the Appendix \ref{RAprompt}.

\section{Improving DPO Training by Data Distribution Adaptation Method}
Since multiple sub-tasks of TOD are executed under limited states,
we conducted DPO training after Supervised Fine-Tuning (SFT) which is more conducive to leveraging the advantages of DPO. However, due to the uncertainty in the distribution of domains in the bad cases, we encountered the degradation issue of DPO mentioned in Section \ref{Back_DPO}. We propose a Data Distribution Adaptation (DDA) method to improve the issue simply and effectively. 


For the first two agents, their results for one real question are all on a specific domain in formatted structures. Therefore, the DPO method is well-suited to leverage its strengths in this scenario. Besides, both of the agents in our method need to complete the complex logical instructions in the prompt, which faces challenges on lightweight LLMs. The DPO method can further improve the weaknesses in training on these instructions during the SFT phase.

When we directly leverage the DPO method to train on the bad cases in the validation set, we also encountered the issue of model degradation after DPO training, which is mentioned in Section \ref{Back_DPO}. We analyze the bad cases and find that, compared to the SFT training data, the rejected data used by DPO had a very uneven distribution in terms of domains. Based on the conclusion that "\emph{the DPO loss function demonstrates a tendency for LLMs to readily learn to avoid generating responses that humans disprefer}" \citep{feng2024towards}, we believe that if the category of the rejected data in the DPO phase is concentrated in a certain category, it will significantly reduce the generation probability for that category after training, which leads to model degradation in that category. Therefore, we generate bad cases for other categories to match the distribution of rejected data across all categories with the data from the SFT phase. In this way, we have effectively alleviated the degradation problem caused by DPO.

\section{Experimental Setup}

\begin{table*}[]
\centering
\resizebox{\textwidth}{!}{
\begin{tabular}{lccccccc}
\toprule
Model                  & BLEU & Inform & Success & Combined & CBE & \#uniq. words & \#uniq. 3-grams \\\midrule
\multicolumn{1}{l}{\emph{\textbf{Traditional model:}}}     & & & & & & &    \\
GALAXY \cite{he2022galaxy}    & 19.6 & 85.4   & 75.7    & 100.2 & 1.75 & 295 & 2275         \\
TOATOD \cite{bang-etal-2023-task}   & 17.0 & 90.0   & 79.8    & 101.9  & - & - & -        \\
Mars-G \cite{sun2023mars}  & \textbf{19.9} & 88.9   & 78.0    & 103.4   &1.65 & 288 & 2264       \\
KRLS \cite{yu2023krls}   & 19.0 & 89.2   & 80.3    & 103.8   & 1.90 & 494 & 3884       \\
DiactTOD \cite{wu2023diacttod}  & 17.5 & 89.5 & 84.2 & 104.4 &2.00 & 418 & 4477  \\ 
SUIT$_{2}$ (DPO-SFT) \cite{kaiser2024learning}  & 16.5 & 90.0 & 87.1 & 105.1 &- & - & -  \\ \midrule
\multicolumn{1}{l}{\emph{\textbf{Large Language Model (LLM):}}}     & & & &    \\
Mistral-7B DARD \cite{gupta2024dard}   & 15.2 & 78.8   & 61.2    & 85.2 & 2.79 & 993 & 13317   \\
Qwen2.5-7B DARD & 14.9  & 80.1   & 61.5    & 85.7 & 2.14 & 902 & 12974 \\ 
SGP-TOD-GPT3.5 \cite{zhang-etal-2023-sgp}    &  9.2 &   82.0 & 72.5 & 86.5 & - & - & - \\
Claude Sonnet 3.0 DARD \cite{gupta2024dard} & 9.5  & \textbf{95.6}   & \textbf{88.0}    & 101.3 & 2.37 & 1197 & 13742 \\ \midrule
\multicolumn{1}{l}{\emph{\textbf{Ours:}}}     & & & & & & &   \\
Qwen2.5-7B DIMF w/o DPO & 14.8  & 90.3   & 75.4    & 97.7 & 2.73 & 1139 & 14305 \\ 
Qwen2.5-7B DIMF & 18.7  & 92.4   & 82.8    & \textbf{106.3} & \textbf{2.81} & \textbf{1231} & \textbf{14328}  \\ 
\bottomrule
\end{tabular}}
\caption{
End-to-end response generation evaluation results on MultiWOZ 2.2 dataset. All results of traditional models are cited from the official leaderboard. We execute the publicly accessible results of the LLM-based model. The "\textbf{bold}" indicates the best score among all the systems of each language pair.
}
\label{maintable}
\end{table*}

\subsection{Dataset \& Evaluation Metrics}
We evaluate our proposed method on the MultiWOZ 2.2 dataset \cite{zang2020multiwoz}.  The dataset is a large-scale multi-domain TOD dataset which contains 10437 conversations and is divided into 
training, validation, and test sets. The dataset comprises 7 domains 
and contains a database for querying the information of a specific domain.

We leverage the traditional evaluation method of the MultiWOZ 2.2 dataset, Inform, Success, and BLEU scores, to evaluate our proposed method. The \textbf{Inform} rate is to check whether the system finds the right entity for the user. The \textbf{Success} rate is to check whether the system provides all the required entity attributes for the user. 
The \textbf{BLEU} measures the fluency compared to the references, which are delexicalized. Finally, the \textbf{Combine} score is a comprehensive metric to indicate the overall performance, which is formulated as: $Combine = \frac{Inform + Success}{2} + BLEU$. Besides, we leverage the Conditional Bigram Entropy (CBE), \#unique words and \#unique 3-grams to evaluate the richness of the response.

\subsection{Baselines \& Setup}
We compare our proposed method with the traditional system and the LLM-based system. We choose several strong baselines fine-tuned on the traditional language models, including GALAXY \cite{he2022galaxy}, TOATOD \cite{bang-etal-2023-task}, Mars-G \cite{sun2023mars}, KRLS \cite{yu2023krls}, DiactTOD \cite{wu2023diacttod}, SUIT \cite{kaiser2024learning}. 
For the LLM-based system, we evaluate the SGP-TOD \cite{zhang-etal-2023-sgp} method which 
builds the TOD system with GPT3.5. Besides, we compare our method with the state-of-the-art LLM-based method, DARD \cite{gupta2024dard}. 
Since the code was not provided of DARD, we independently replicate the results of the DARD method on the Qwen2.5-7B model.

We select Qwen2.5-7B-Instruct \cite{qwen2.5} as our foundation model for our proposed method. The details of our training settings are attached in the Appendix \ref{DDAapp}.


\section{Experiments}

\subsection{Main Results}
We present the results of our proposed DIMF and other baselines in Table \ref{maintable}. Specifically, each agent in DIMF is first fine-tuned on the entire training set under supervision and then trained using the DPO method on the validation set. The results show that our proposed method achieves the best Combined score among all the baselines.

\begin{table*}[]
\resizebox{\textwidth}{!}{
\begin{tabular}{lccccccccccccccc}
\toprule
\multicolumn{1}{c}{\multirow{2}{*}{Model}}  & \multicolumn{3}{c}{Attraction} & \multicolumn{3}{c}{Hotel}  & \multicolumn{3}{c}{Restaurant}   & \multicolumn{3}{c}{Taxi} & \multicolumn{3}{c}{Train} \\ \cmidrule(r){2-4} \cmidrule(r){5-7} \cmidrule(r){8-10}  \cmidrule(r){11-13}  \cmidrule(r){14-16} 
\multicolumn{1}{c}{}         & BLEU & Info. & Succ. &  BLEU & Info. & Succ.  &  BLEU & Info. & Succ.  &  BLEU & Info. & Succ. &  BLEU & Info. & Succ.\\  \midrule
\multicolumn{16}{l}{\emph{\textbf{Base System (All agents trained with SFT)}}}    \\
DIMF-base & 14.8 & 98.7 & 83.2 & 14.2 & 89.6 & 74.8 & 13.7 & 96.2 & 85.3 & 15.2 & 100.0 & 85.1 & 15.0 & 90.1 & 78.1 \\ \midrule
\multicolumn{16}{l}{\emph{\textbf{w/ Intent Classification Agent DPO}}}    \\
DPO-Ori & \colorbox{mygray}{11.9} & \colorbox{mygray}{86.3} & \colorbox{mygray}{71.0} & 13.1 & 90.0 & 75.2 & 12.2 & \colorbox{mygray}{90.2} & 79.1 & \colorbox{mygray}{12.7} & 100.0 & \colorbox{mygray}{73.3} & 15.0 & 90.5 & 80.0\\
DPO-DDA & 14.8 & 99.1 & 83.7 & 13.7 & 90.3 & 76.7 & 13.6 & 96.2 & 85.3 & 15.6 & 100.0 & 86.0 & 14.9 & 91.4 & 78.4\\ \midrule
\multicolumn{16}{l}{\emph{\textbf{w/ Intent Classification Agent DPO-DDA \& Slot Filling Agent DPO}}}    \\
DPO-Ori & \colorbox{mygray}{11.0} & \colorbox{mygray}{81.7} & \colorbox{mygray}{69.4} & 12.7 & 80.5 & 73.1 & 12.9 & \colorbox{mygray}{83.4} & \colorbox{mygray}{73.3} & 14.8 & 100.0 & 79.1 & \colorbox{mygray}{12.5} & \colorbox{mygray}{79.6} & \colorbox{mygray}{71.9}\\
DPO-DDA & 17.1 & 99.1 & 90.2 & 16.2& 90.6 & 83.6 & 15.9 & 96.2 & 89.7 & 17.1 & 100.0 & 88.2 & 16.7 & 90.8 & 83.2\\ \midrule
\multicolumn{16}{l}{\emph{\textbf{w/ Intent Classification Agent DPO-DDA \& Slot Filling Agent DPO-DDA \& Response Agent DPO}}}    \\
DPO-Ori & 19.6 & 99.1 & 90.2 & 17.3 & 91.0 & 83.1 & 16.0 & 96.2 & 89.0 & 18.8 & 100.0 & 89.6 & 19.2 & 92.3 & 82.7 \\
DPO-DDA & 19.4 & 99.1 & 90.2 & 17.7& 91.3 & 84.0 & 16.3 & 96.5 & 89.7 & 18.6 & 100.0 & 89.6 & 19.5 & 92.3 & 83.2
\\ \bottomrule
\end{tabular}}
\caption{Results of different DPO training method on each agent of DIMF. The \colorbox{mygray}{gray} data indicates the degradation data. The DPO-Ori represents the original DPO training method which directly leverage the bad cases for training. The DPO-DDA represents our proposed Data Distribution Adaptation method. }
\label{DDAtable}
\end{table*}

Compared with the traditional models, DIMF has become more powerful in slot extraction which corresponds to the scores of Inform and Success. This also demonstrates that the method of separating the complex tasks in our DIMF can effectively enhance the system's capability. 
As for the Large Language Model, our model has outperformed the same size model on all evaluation metrics. The results of the DARD method on the Qwen model prove the advancement of our method. Besides, compared to the large-scale LLMs, our method has a significant improvement on the BLEU. Moreover, unlike the DARD method, we use a single model for all domains 
which demonstrates a better generalization of our method.

The last three metrics evaluate the textual richness of the model response. The results show that our method significantly outperformed other models. This also demonstrates the advantages of LLMs compared to the traditional models: the diversity of responses can provide users with a better interactive experience in real-world scenarios.


\subsection{Results of Data Distribution Adaptation Method for DPO Training}
\label{DDAtest}

In this section, we aim to demonstrate that our Data Distribution Adaptation method can effectively mitigate the issue of DPO degradation. The test set contains 5 domains with different numbers (Attraction (396), Hotel (394), Restaurant (437), Taxi (195) and Train (495)). We present the results of each domain in Table \ref{DDAtable}. We define that if the performance of a specific domain drops below the average accuracy, then the model has a degradation issue in that domain. Due to testing issues, the Inform for the Taxi did not change. The distribution of bad cases on the test set is similar to the validation set, so we will directly analyze the results on the test set between the two DPO methods.

\begin{figure}[t]  
\centering
\includegraphics[width=0.45\textwidth]{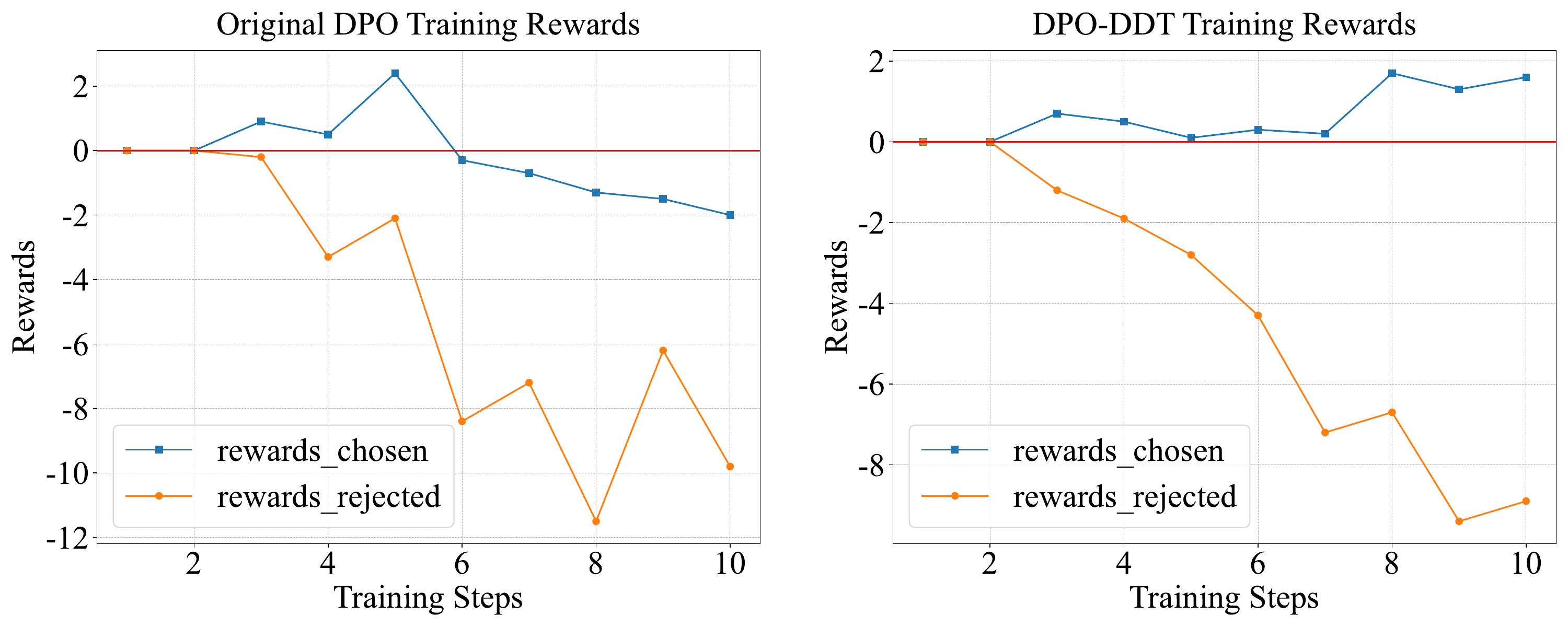} 
\caption{The rewards of the chosen data and rejected data during the Slot Filling Agent DPO training. The left figure is the original DPO method and the right one is our proposed DDA method. The red line represents the reward of 0.} 
\label{DPOlossfig}
\end{figure}

\begin{figure*}[htbp]  
\centering
\includegraphics[width=\textwidth]{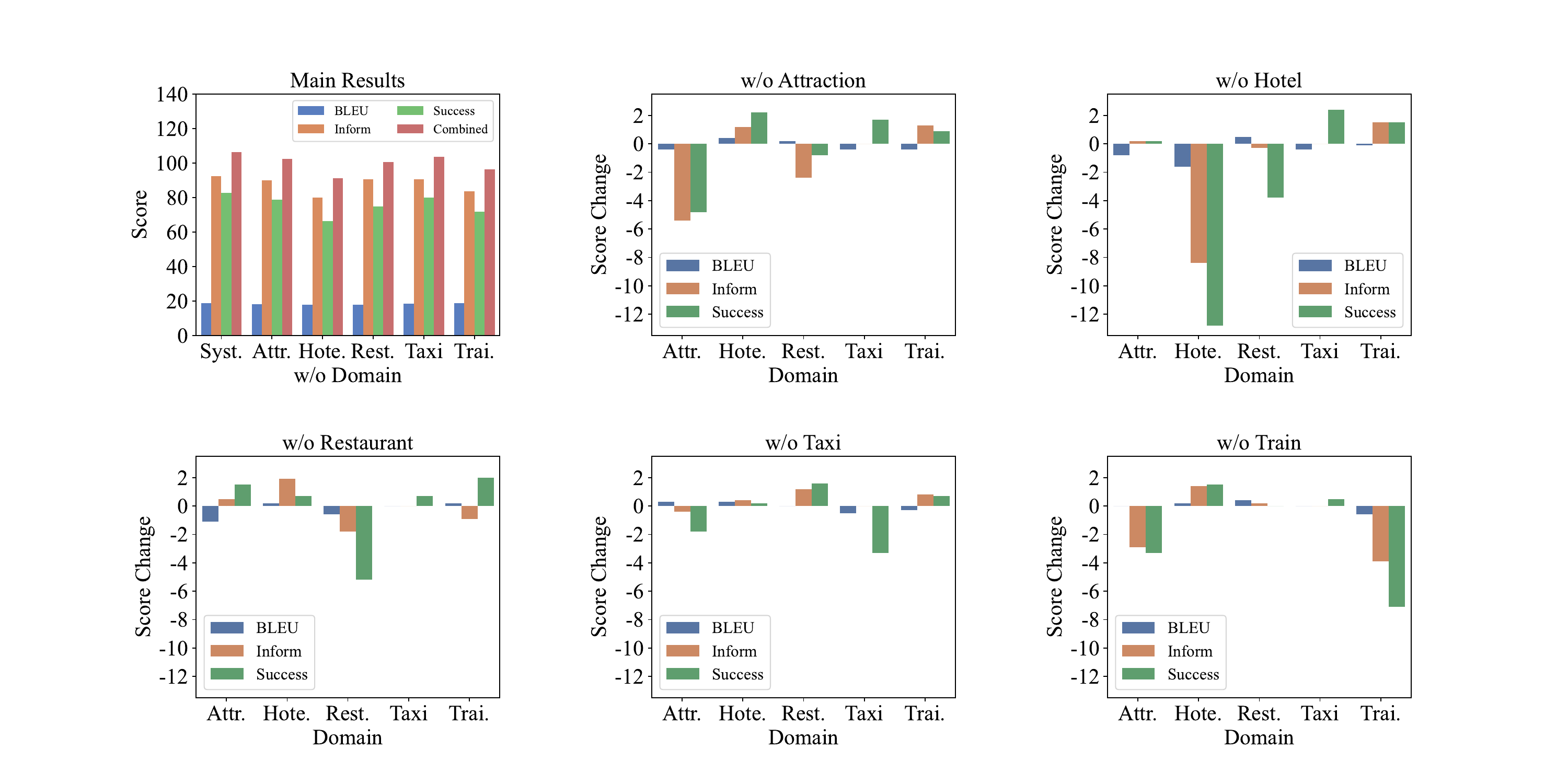} 
\caption{The Results of the DIMF after removing training data from a specific domain. The first sub-figure shows the results of the system after removing different domains. The other sub-figures shows the performance of each domain after removing a specific domain respectively.} 
\label{zerofig}
\end{figure*}

\noindent\textbf{Intent Classification Agent: } Most of the errors are concentrated in the Hotel and Train domains after SFT training. Therefore, these two domains tend to appear more frequently in the chosen data of the original DPO method. Most of the data in the rejected data set belongs to the other three domains. The results show that the data distribution on the rejected data of the original DPO training method leads to the decrease on these three domains.

\noindent\textbf{Slot Filling Agent: } 
During the DPO training phase of Slot Filling Agent, the degradation issue appeared in more domains. We find that many bad cases at this stage occurred when information from multiple rounds of dialogue needed to be inherited. These bad cases were very unevenly distributed across different slot categories, such as area, leading to degradation in various domains. 

\noindent\textbf{Response Agent: } The degradation issue of DPO is not significant in Response Agent. 

\noindent\textbf{Training Rewards: } We show the training rewards of the chosen data and rejected data during the DPO training process of the Slot Filling Agent in Figure \ref{DPOlossfig}. In an ideal situation, "reward\_chosen" should be greater than 0 and increase as training progresses, while "reward\_rejected" should be less than 0 and decline. As we can see, the original DPO method encountered issues with the chosen reward decreasing and becoming less than 0. This issue leads to the degradation of the DPO training process, which demonstrates our analysis above. Our proposed DDA method can efficiently address this problem which is shown in the right figure. The experimental results demonstrate the effectiveness of our DDA-based DPO method. The other agents' results are appended in Appendix \ref{apploss}.

\subsection{Zero-shot Evaluation}

We evaluate the zero-shot capabilities of our proposed framework in this section. For each agent in our method, we remove the data of one domain during the training process. We show the performance of the total system and each domain after removing the specific domain in Figure \ref{zerofig}.

The first sub-figure presents the results of the system. The x-axis represents the results of the original system and the results after removing the training data of different domains. The results indicate that, except for the Hotel and Train domains, the performance of the system does not have a significant decrease compared to the original system after removing other domains. 
As for the Hotel and Train, the results in Table \ref{DDAtable} show that these two domains are more challenging, and our system performs relatively poorly on them. We believe this is the reason for the decline of performance. Nevertheless, the performance of our proposed method still exceeds the same size LLM in Table \ref{maintable} in these two experiments. The result demonstrates that our method enhances the generalization ability of the TOD system by refining tasks within the system.

The other sub-figures present the results on each domain after removing different domains. The results indicate that the accuracy of the specific domain decreased after removing its corresponding data, particularly in the Hotel and Train domains, which confirms the analysis in the last paragraph. Besides, we also observed a phenomenon in the experiment that the performance of some other domains declined after removing one domain. We think that this may be caused by the reduction in data diversity. Moreover, we find that the zero-shot setting has little impact on the BLEU metric.

\begin{table}[]
\centering
\resizebox{0.48\textwidth}{!}{
\begin{tabular}{lcccc}
\toprule
Model   & BLEU & Inform & Success & Combined\\\midrule
Qwen2.5-7B Single Agent   & 10.3 & 59.8  & 37.4 & 58.9 \\
Qwen2.5-7B Two Agents & 14.9 & 80.1 & 61.5  & 85.7 \\
Qwen2.5-7B DIMF w/o DPO & 14.8  & 90.3   & 75.4    & 97.7 \\
\bottomrule
\end{tabular}}
\caption{
Ablation studies results on our proposed DIMF. We compare the performance between different number of agents trained with SFT method.
}
\label{framework_table}
\end{table}

\begin{figure*}[htbp]  
\centering
\includegraphics[width=\textwidth]{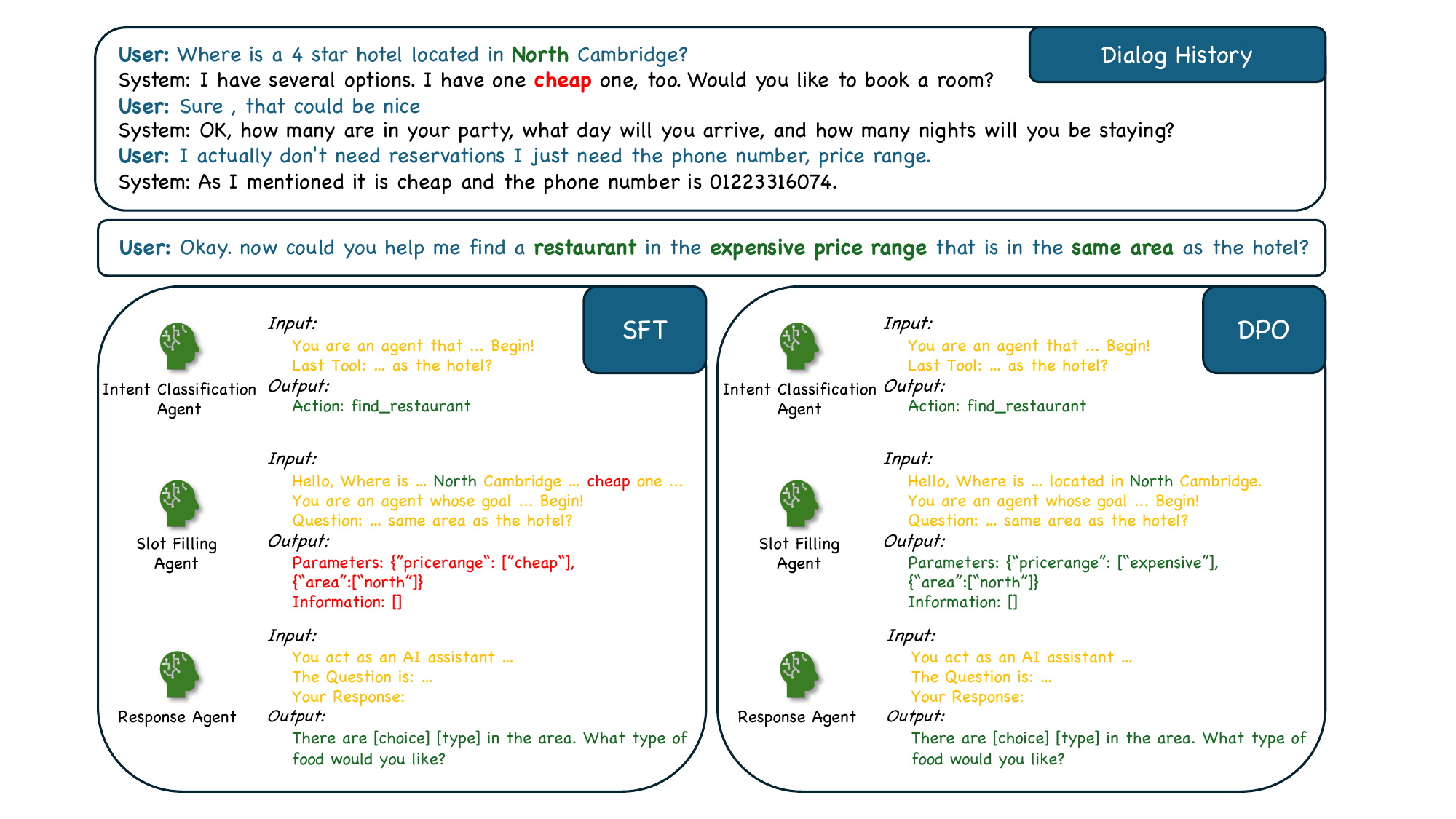} 
\caption{An example of one round of the conversation between user and our DIMF. This case contains the history of the conversation, the question of the user and the generation process of DIMF trained with different methods. The red word represents incorrect information and responses, and green represents correct ones.} 
\label{casefig}
\end{figure*}

\subsection{Ablation Studies}
\subsubsection{Ablation Studies on Framework}
In this section, we evaluate different frameworks to demonstrate the advantage of our DIMF. Specifically, we combine all the training data of our proposed three agents to train a single agent for TOD task. Besides, we combine the intent classification and slot filling agents into a single agent to train a two-agents system. All the frameworks are trained with SFT method. As shown in Table \ref{framework_table}, the DIMF brings a significant improvement for the system, especially on the Inform and Success metrics, which demonstrates the better accuracy of our DIMF.

\begin{table}[]
\centering
\resizebox{0.48\textwidth}{!}{
\begin{tabular}{lcccc}
\toprule
Model   & BLEU & Inform & Success & Combined\\\midrule
Qwen2.5-7B DIMF    & 18.7  & 92.4   & 82.8    & 106.3     \\
w/o R. DPO & 16.8 & 91.2 & 81.3  & 103.1 \\
w/o R. \& S. DPO  & 14.6 & 91.2  & 76.8 & 98.6 \\
w/o R. \& S. \& I. DPO & 14.8 & 90.3 & 75.4  & 97.7 \\
\bottomrule
\end{tabular}}
\caption{
Ablation studies results on our proposed DDA-based DPO method. The R., S. and I. represent Response Agent, Slot Filling Agent and Intent Classification Agent separately. Each row in the table is based on the last row with the DPO method removed. 
}
\label{dpotable}
\end{table}

\subsubsection{Ablation Studies on DPO}
\label{abtest}

In order to better understand the effect of the DPO training method on each agent 
, we perform an ablation test and present the results in Table \ref{dpotable}. All the results in this section are obtained using our proposed DDA training strategy for DPO. The results show that DPO training improves the accuracy of each stage in the system, thereby alleviating the problem of error accumulation.

As we can see in Table \ref{dpotable}, compared to the other two agents, the improvement of DPO in the Intent Classification Agent is limited. We believe this is because the model trained after SFT already possesses relatively good capabilities. However, the Slot Filling Agent and the Response Agent still show significant improvement in the BLEU and Success metrics after our DDA-based DPO training. The experimental results also demonstrate that, compared to other methods, our DIMF approach, which trains the Slot Filling Agent separately and isolates the Response Agent, is very effective in enhancing performance in the TOD system.

\subsection{Case Study}
\label{casestudy}

To further understand the detailed process of our method, we provide a case study that contains the output of each agent for a specific user's question. We select a more challenging case that requires inheriting information from the historical dialogue.

As shown in Figure \ref{casefig}, when our system receives a user's question, the question first be directly transferred into the Intent Classification Agent without dialogue history to obtain the user's intent. Next, the slot prompt of this specific domain with the dialogue history is input into the Slot Filling Agent to obtain the specific information in this domain that the user needs to inquire about. Finally, the results queried from the database 
are input into the Response Agent to obtain the response for the user.

In this case, we can see that the user does not specify the specific information in the "area" slot directly. The system needs to inherit this information and remove another irrelevant slot "cheap" from the last intent.
The Slot Filling Agent implements this ability by adding the logic rule about inheriting historical dialogue information in the prompt. However, as shown in this case, the lightweight LLMs trained with the SFT method cannot fully learn this capability and sometimes make mistakes on this issue. The DPO method provides targeted training for this capability, effectively improving the shortcomings of the SFT method and improving the system's performance.

\section{Conclusion}
In this work, we propose a new framework, Domain-Independent Multi-Agent Framework (DIMF), for TOD systems. We separate the original complex task into three sub-tasks, Intent Classification Agent, Slot Filling Agent, and Response Agent, which reduces the complexity of each agent and makes the performance of lightweight LLMs more reliable.
Our framework trained on the Qwen2.5-7B achieves better performance compared with
all the baselines. Besides, during the training process, we leverage the advantages of the DPO method on this task to address the deficiencies in understanding logical rules in prompts during the SFT process. We propose a Data Distribution Adaptation (DDA) method to mitigate the degradation issues of DPO. The results prove that our method is easy to implement and effective. Moreover, we demonstrate that our system can better utilize the generalization capabilities of LLMs and has a good zero-shot ability.

\section{Limitations}
In this work, with a carefully designed TOD framework, we have revealed that current systems on TOD tasks severely suffer from insufficient task independence and model scalability. We further propose the DIMF and DDA training methods to mitigate the phenomenon. However, our work still has limitations. Firstly, during the tool invocation stage, we directly access the database based on the results of the Slot Filling Agent. When facing more diverse, complex, or real tools, it may be necessary for the model to generate a unified invocation statement to address this issue. 
Secondly, our current reinforcement learning method mainly leverages the improved DPO method. Nowadays, the Group Relative Policy Optimization (GRPO) \cite{shao2024deepseekmath} shows impressive performance, we will apply this new method on our framework in our future work.

\bibliography{custom}

\appendix

\section{Prompt}

\subsection{Prompt of Intent Classification Agent}
\label{IAprompt}

We show an example of the Intent Classification Agent at the second-round of the conversation in Table \ref{IAprompt}.

\begin{table*}[]
\caption{
Intent Classification Agent prompt
}
\label{IAprompttable}
\resizebox{\textwidth}{!}{
\begin{tabular}{l}
\toprule
You are an agent that helps users choose the right tool or tools from the given tools list to solve their problems. \\
\\
For each tool, you are first given its description and required parameters. Then, a logic module specifically explains the \\
logical information needed for this tool to handle multi-turn conversation issues. \\
\\
\#\# Tool APIs \\
\\
find\_hotel: search for a hotel to stay in \\
book\_hotel: book a hotel to stay in \\
find\_train: search for trains that take you places \\
book\_train: book train tickets \\
find\_attraction: search for places to see for leisure \\
find\_restaurant: search for places to wine and dine \\
book\_restaurant: book a table at a restaurant \\
find\_hospital: search for a medical facility or a doctor \\
find\_taxi: find or book taxis to travel between places \\
find\_bus: search for a bus \\
find\_police: search for police station \\
other: This tool is used to handle problems that cannot be addressed by any other tools. \\
\\
\#\# Task Logic \\
If last query is find\_restaurant, the user can use the same tool for the following types of query:\\
  - restaurant-pricerange: price budget for the restaurant. only allowed values: [cheap, expensive, moderate] \\
  - restaurant-area: area or place of the restaurant. only allowed values: [centre, east, north, south, west] \\
  - restaurant-food: the cuisine of the restaurant you are looking for. \\
  - restaurant-name: name of the restaurant. \\
  - restaurant-bookday: day of the restaurant booking. only allowed values: \\ 
    \qquad[monday, tuesday, wednesday, thursday, friday, saturday, sunday] \\
  - restaurant-bookpeople: how many people for the restaurant reservation. only allowed values: [1, 2, 3, 4, 5, 6, 7, 8] \\
  - restaurant-booktime: time of the restaurant booking.  \\
\\
\#\# Output Format \\
\\
Use the following format: \\
\\
Last Tool: the tool used in last query \\
Question: the input question you must answer \\
Action: the action to take \\
Finish! \\
\\
Begin! \\
\\
Last Tool: find\_restaurant \\
Question: Any sort of food would be fine. Could I get the phone number for your recommendation? \\
\bottomrule
\end{tabular}}
\end{table*}

\subsection{Prompt of Slot Filling Agent}
\label{SAprompt}

We show an example of the Slot Filling Agent of the restaurant domain at the second-round of the conversation in Table \ref{SAprompt}.

\begin{table*}[]
\caption{
Slot Filling Agent Filling prompt
}
\label{SAprompttable}
\resizebox{\textwidth}{!}{
\begin{tabular}{l}
\toprule
You are an agent whose goal is to extract the required tool parameters and the content the user wants to query from their questions. \\
\\
For a specific query, you are first given the parameters corresponding to the restaurant tool. Besides, you have also been informed the information \\
that the specific information this tool can query. Finally, you are given the logic distinguish between Tool Parameters and Tool Information. \\
\\
\#\# Tool Parameters\\
\\
restaurant-pricerange: price budget for the restaurant. only allowed values: [cheap, expensive, moderate]\\
restaurant-area: area or place of the restaurant. only allowed values: [centre, east, north, south, west]\\
restaurant-food: the cuisine of the restaurant you are looking for. \\
restaurant-name: name of the restaurant. \\
restaurant-bookday: day of the restaurant booking. only allowed values: [monday, tuesday, wednesday, thursday, friday, saturday, sunday]\\
restaurant-bookpeople: how many people for the restaurant reservation. only allowed values: [1, 2, 3, 4, 5, 6, 7, 8]\\
restaurant-booktime: time of the restaurant booking. \\
\\
\#\# Tool Information\\
\\
The user can use restaurant tool to query the following questions:\\
address: the address of the restaurant. \\
area: the location information of the restaurant can be selected from the following options: [east, south, west, north]. \\
food: the food of the restaurant. \\
id: the id number of the restaurant. \\
introduction: the introduction of the restaurant. \\
location: the coordinates of the restaurant. \\
name: the name of the restaurant. \\
phone: the phone of the. \\
postcode: the postcode of the restaurant. \\
pricerange: the level of the price of the restaurant. \\
type: . \\
\\
\#\# Task Logic\\
\\
 - If the user's question includes a slot name and the slot value, then this query information \\ \quad belongs to the tool Parameters, and output must in a JSON type. \\
 - If the user's question only includes a slot name without value, then this query information belongs to the tool Information.\\
 - If the user needs information from the historical conversation, you can obtain it from the History Conversation slot. \\
\\
\#\# History Conversation slot\\
\\
restaurant:\\
{"area": ["centre"], "pricerange": ["expensive"]}\\
\\
\#\# Output Format \\
\\
Use the following format:\\
\\
Question: the input question you must answer\\
Action: the tool that user used\\
Parameters: must a JSON object of the slot with its value\\
Information: the tool information in a list object\\
Finish!\\
\\
Begin!\\
\\
Question: Any sort of food would be fine, as long as it is a bit expensive. Could I get the phone number for your recommendation?\\
Action: restaurant\\
\bottomrule
\end{tabular}}
\end{table*}

\subsection{Prompt of Response Agent}
\label{RAprompt}

We show an example of the Response Agent in Table \ref{RAprompt}.

\begin{table*}[]
\caption{
Response Agent prompt
}
\label{RAprompttable}
\resizebox{\textwidth}{!}{
\begin{tabular}{l}
\toprule
You act as an AI assistant to reponse user's question relied some given informations.\\
You should always communicate with the user in the first person and respond in a personified manner.\\
The Question is: I need train reservations from norwich to cambridge \\
\\
\#\# Responce Rules\\
\\
You should respond according to the following rules:\\
\\
Make a conclusion based on the the user's question, Observation and conversation history.
If there are several options, \\
you can first respond the total number of the option, make a conclusion of the "conclusion informations" and then ask the \\ question about the informations in "question content" \\
  - example: "I have xxx options matching your request. Waht's the xxx you want to xxx"\\
  - example with conclusion informations: "I have xxx options matching your request. The range of xxx in these options is xxx. \\ Waht's the xxx you want to xxx"\\
If there is only one options, you can make a conclusion if it and respond to the user.\\
All the specific information in the response should be in this format: [type\_name]\\
\\
\#\# Observation\\
\\
train information:\\
option number: 133\\
question content: arriveby, leaveat, trainid, day, price\\
conclusion informations:\\
 arriveby: 06:35, 07:35, 08:35, 09:35, 21:35, 22:35, 23:35, 24:35\\
 leaveat: 05:16, 06:16, 07:16, 08:16, 20:16, 21:16, 22:16, 23:16\\
\\
\\
\#\# Note\\
\\
You should respond with more varied expressions. \\
Your respond should contain all the information in Observation, and your reply should no more than 25 words.\\
\\
Your Response:\\
\bottomrule
\end{tabular}}
\end{table*}

\section{DDA Data Generating Method}
\label{DDAapp}

We generate the training dataset tailored to each agent for SFT method based on the MultiWOZ 2.2 dataset. For the DDA method, the data-generating method is as follows: 

We first introduce the preference pairs implementation method:
\begin{itemize}
    \item Positive samples: Responses with correct intent/slot predictions. As for the Response Agent, we select good cases based on a certain threshold of BLEU.
    \item Negative samples: Responses with incorrect predictions and under the threshold.
\end{itemize}

To conduct the DDA method, our negative example sampling strategies for distribution balancing are:
\begin{itemize}
    \item Intent Classification Agent: We randomly replace target intents with incorrect ones.
    \item Slot Filling Agent: We either replace slot values with other values from the dialogue context or remove values from multi-value slots.
    \item Response Generation Agent: We modify response rules to generate contextually inappropriate responses.
\end{itemize}

All the agents are fully fine-tuned and conducted on 8 A100 GPUs with 40GB of RAM for 2 epochs.

\section{DPO Training Loss}
\label{apploss}

\begin{figure}[htbp]  
\centering
\includegraphics[width=0.45\textwidth]{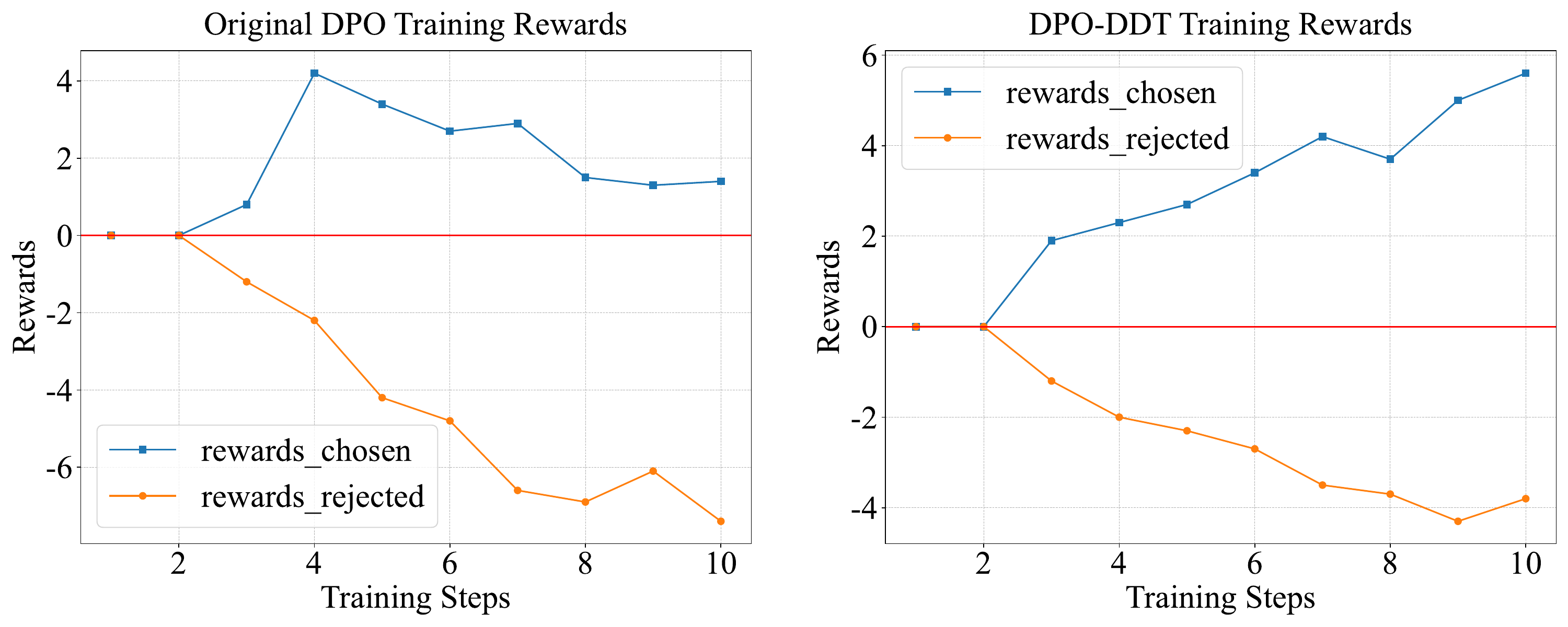} 
\caption{The rewards of the chosen data and rejected data during the Intent Classification Agent DPO training. } 
\label{I_loss_fig}
\end{figure}

\begin{figure}[htbp]  
\centering
\includegraphics[width=0.45\textwidth]{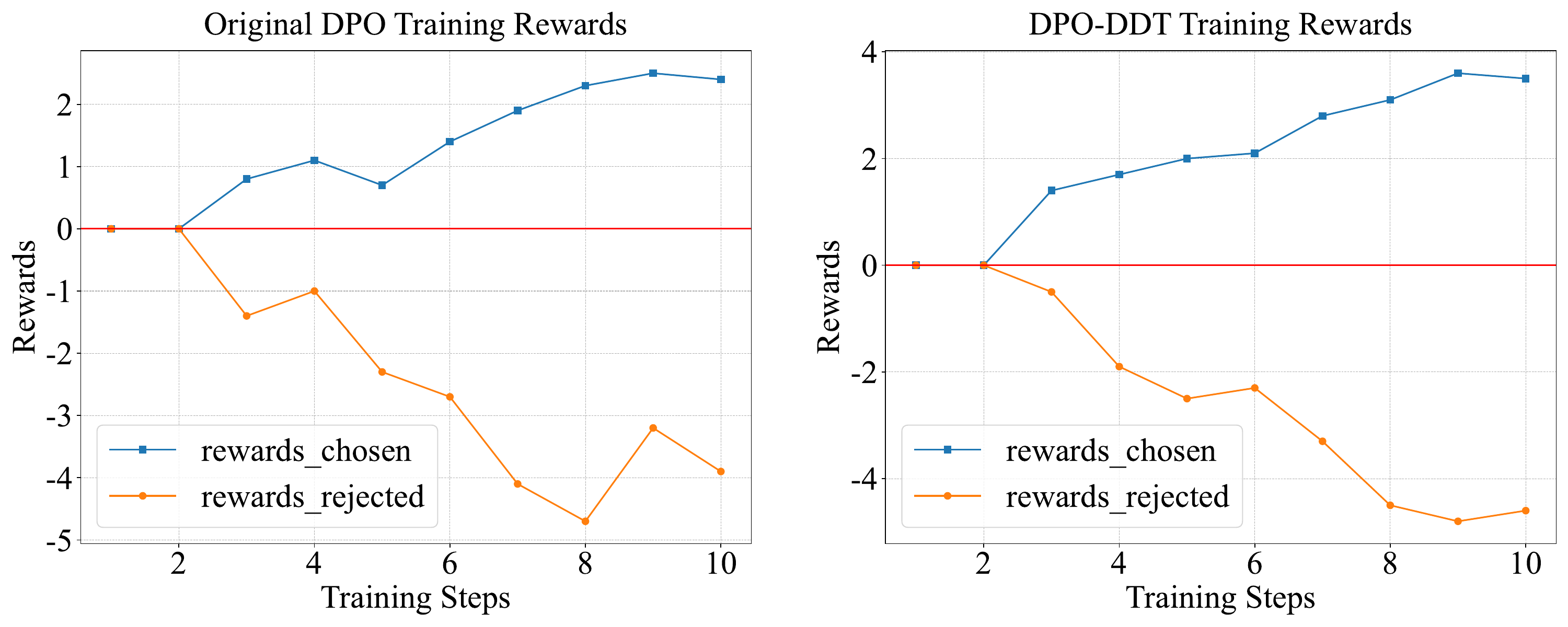} 
\caption{The rewards of the chosen data and rejected data during the Response Agent DPO training. } 
\label{R_loss_fig}
\end{figure}

We present the results of the reward loss of the Intent Classification Agent and Response Agent in Figure \ref{I_loss_fig} and Figure \ref{R_loss_fig}. Compared to Slot Filling Agent, the degradation issues on the original DPO method are not as severe for these two models. The Intent Classification Agent experienced a reduction in chosen reward, while the training of the Response Agent was relatively normal.

\end{document}